\documentclass{jpsj2}

\title{Microwave-Absorption-Induced Heating of Surface State Electrons on Liquid $^3$He}

\author{\textsc{Denis Konstantinov}$^{1}$\thanks{E-mail address: konstantinov@riken.jp}, \textsc{Hanako Isshiki}$^{1,2}$,  \textsc{Hikota Akimoto}$^{1}$, \textsc{Keiya Shirahama}$^{2}$, \textsc{Yuriy Monarkha}$^{1,3}$, and \textsc{Kimitoshi Kono}$^{1}$}

\inst{$^{1}$Low Temperature Physics Laboratory, RIKEN, Hirosawa 2-1, Wako 351-0198, Japan\\
      $^{2}$Department of Physics, Keio University, Hiyoshi 3-14-1, Yokohama 223-8522, Japan\\
      $^{3}$Institute of Low Temperature Physics and Engineering, 47 Lenin Avenue, 61103 Kharkov, Ukrain}

\abst{
A resonance-induced change in the resistivity of the surface state electrons (SSE) exposed to the microwave (MW) radiation is observed. The MW frequency corresponds to the transition energy between two lowest Rydberg energy levels. All measurements are done with electrons over liquid $^3$He in a temperature range 0.45$-$0.65~K, in which the electron relaxation time and the MW absorption linewidth are determined by collisions with helium vapor atoms. The input MW power is varied by two orders of magnitude, and the resistivity is always found to increase. This effect is attributed to the heating of electrons with the resonance MW radiation. The temperature and the momentum relaxation rate of the hot electrons are calculated as a function of the MW power in the cell, and the Rabi frequency is determined from the comparison of the theoretical result with the experiment. In addition, the broadening of the absorption signal caused by the heating is studied experimentally, and the results are found to be in good agreement with our calculations.
}
\kword{surface state electrons on liquid helium, Rydberg states, microwave absorption, electron conductivity, hot electrons}

\begin{document}

\maketitle

\section{Introduction}

$\indent$It is well known that an electron outside the liquid is attracted to helium surface by a weak image force, 
while it is prevented to enter the liquid by a strong repulsion due to the Pauli exclusion principle. As a result, the SSE 
are trapped into potential well and their motion in the direction normal to the surface is restricted to the 
hydrogen-like bound states with energies $\epsilon_n=-\Delta/n^2$, where $n\geq 1$ is an integer number and $\Delta$ is the effective Rydberg energy. The latter is about 4~K for electrons over $^3$He, and at typical experimental conditions $T\lesssim 1$~K almost all electrons are in the ground level. In each Rydberg level, the electrons are free to move along the surface and thus form a series of two-dimensional conductive subbands. At $T\gtrsim 0.5$~K, the intra-subband relaxation times and the inter-subband transition linewidths are determined by the interaction of the electrons with helium vapor atoms, while at $T\lesssim 0.3$~K they are limited by the interaction with ripplons because the vapor density becomes negligible. In both cases, the momentum relaxation rate of electrons differs for electrons occupying different subbands. 

The inter-subband transitions of the SSE above $^4$He were first directly observed by Grimes \textit{et al}~\cite{Grimes1} 
who measured the temperature dependence of linewidth for transition between the the ground level and the first excited Rydberg level in the vapor-atom scattering regime. Recently Collin \textit{et al}~\cite{Collin} extended these measurements to lower temperatures to cover the ripplon scattering regime and found a good agreement with theoretical predictions~\cite{Ando}. In these experiments the electrons are Stark tuned in resonance with MW field by varying the vertical electric field exerted on electrons. The MW absorption due to electron transitions from the ground level to the excited levels is then detected as the variation of the power of the MW radiation passing through the cell. Recently, we performed a similar experiment~\cite{Isshiki} to directly observe the transitions to the excited states for electrons above $^3$He. The temperature dependence of the transition linewidth for the first excited state was found also in agreement with the theory~\cite{Ando}, although the absolute value of the linewidth was somewhat higher than the theoretical estimate.

The direct detection of MW power absorption is not the only way to observe the inter-subband transitions of the SSE. Several experiments showed the changes in the mobility for the electron motion parallel to the surface induced by the resonant microwave absorption. First, Edel'man~\cite{Edel'man1} observed the increase in the resonant cyclotron absorption when the SSE were exposed to MW radiation, which caused the electron transitions from the ground to the excited subbands. Later, Volodin and Edel'man~\cite{Volodin} observed the resonant change in the electron conductivity as a result of the interaction with microwaves. Both experiments were done with electrons above $^4$He and $^3$He and in the temperature range 0.3$-$0.5~K, which corresponds to ripplon scattering and vapor-atom scattering regimes for $^4$He and $^3$He respectively. The authors immediately pointed out that the observed change in conductivity is most likely due to heating of electrons with absorbed MW power. They found out that in a case of $^4$He the conductivity increases, which should be expected for hot electrons whose momentum relaxation time is limited by ripplon scattering~\cite{Saitoh_hot}. In a case of $^3$He however a rather complicated behaviour was observed, namely the conductivity decreased at small power of MW field, and it increased at higher power. As far as we know, no satisfactory explanation for this complicated behaviour was given yet.

Below, we describe an experiment in which the microwave-resonance-induced change in the resistivity of the SSE above liquid $^3$He is investigated in the temperature range from 0.45 to 0.65~K. Contrary to the result obtained by Volodin and Edel'man~\cite{Volodin}, we find that the conductivity decreases at the resonance for all values of the input MW power, which was varied by more than two orders of magnitude in our experiment. We explain the observed change in the resistivity by the heating of the SSE under the conditions of the MW resonance, and propose a theory which adequately describes the heating process and its effect on the electron scattering rate. In our earlier publication,~\cite{PRL} we reported the first results of our resistivity measurements and illustrated the heating mechanism using a rather simple two-level model. In this paper, we give a detailed description of the experimental procedure and the obtained results, and perform a thorough analysis of the processes of MW absorption and the heating of SSE taking a full account of the thermal population of the higher Rydberg states. In addition, we investigate the effect of the heating on the MW absorption linewidth and make comparison of the experimental results with the predictions of our theory.

The paper is organized as follows: in Sec.~2, a brief mention of previous experimental and theoretical studies of hot electrons is given and the important notations for the electron scattering and the energy relaxation rates are introduced. In Sec.~3, we give the description of the experimental setup and the measurement procedure, and present the results of our resistivity measurements. In this section we also present our hot electron theory: describe the heating mechanism, calculate the electron temperature and the electron scattering rate as a function of microwave power, and make the comparison with the experimental results. The effect of SSE heating on the absorption linewidth is discussed in Sec.~4. In the last section, some conclusions are drown and the future plans are outlined. 

\section{Theoretical background}

It has been realized that due to the high mobility and the slow energy relaxation rate the SSE can be easily heated up to the temperatures much higher than the temperature of the liquid substrate. If the electron temperature is sufficiently high to cause an appreciable thermal population of the higher excited subbands, the electrons are said to be "hot". The occupation of the higher subbands has an important consequence: since the electron scattering rate depends on the subband index, the energy and the momentum relaxation rates, which are determined by the collisions of electrons with scatterers, depend strongly on the electron temperature and on the degree of higher subband population. This property makes the electron heating to be easily observable. In addition to the experiments mentioned in the previous section, the hot electron effect was seen in the transport~\cite{Bridges}, the plasmon resonance~\cite{Grimes2} and the cyclotron-resonance measurements~\cite{Edel'man2}. Most recently, an interesting effect, at least partially attributed to the electron heating, was reported by Penning \textit{et al} who observed a large change of magneto-conductivity of SSE under the cyclotron-resonance conditions.~\cite{Penning} 

The first theoretical studies of hot surface electrons were done by Crandall~\cite{Crandall} who took into account the occupation of higher subbands but ignored inter-subband scattering, and by Shikin~\cite{Shikin} who considered the population of the quasi-continuous spectrum by electrons heated under the cyclotron-resonance conditions. A rather complete theoretical treatment of the problem was given by Saitoh and Aoki~\cite{Saitoh_hot} for the SSE heated by the low-frequency in-plane driving field, and by Aoki and Saitoh~\cite{Aoki} for the SSE heated by the high-frequency cyclotron-resonance field. In Ref.~7, authors employed an effective temperature approximation and calculated the electron temperature as a function of the driving field by solving the energy balance equation. This allowed them to calculate the electron scattering rate as a function of the driving field for both vapor-atom and ripplon scattering. An important result obtained by Saitoh and Aoki was that as the electron temperature goes up and electrons become distributed over many subbands, the inter-subband scattering events become particularly important, and that this leads to an increase of the electron scattering rate in the vapor-atom scattering regime.  

Keeping in mind the temperature range of our experiment, in this paper we will consider only the case of the scattering by the vapor atoms. The electron motion perpendicular to the helium surface is described by the Schrodinger equation

\begin{equation}
\Biggl [ -\frac{\hbar^2}{2m_{\textrm{e}}} \frac{\textrm{d}^2}{\textrm{d}z^2} - \frac{(\varepsilon -1)e^2}{4(\epsilon +1)z} +eE_{\perp}z \Biggr ] \psi_n(z)=\epsilon_n \psi_n(z),
\label{eq:wave}
\end{equation}

\noindent where $z>0$ is the distance from the surface, $\psi_n$ and $\epsilon_n$ are the eigenfunctions and eigenvalues of the above equation corresponding to the quantum number $n$, $m_{\textrm{e}}$ is the electron mass and $-e$($<$0) is the electron charge, $\varepsilon$ is the dielectric constant of helium, $E_{\perp}$ is the electric field applied perpendicular to the surface. Here we assume an infinitely large surface barrier potential, and the above equation should be solved with the boundary condition $\psi_n(0)=0$. For further discussion, it is convenient to introduce a number of notations similar to those used in Ref.~7. First, let us define the electron scattering rates $\nu_{mn}$ from the subband with index $m$ to the subband with index $n$, which for $m\geq n$ are given by

\begin{equation}
\nu_{mn}=\frac{\pi \hbar N_{\textrm{G}} A}{m_{\textrm{e}} B_{mn}} ,
\label{eq:nu}
\end{equation}

\noindent where $N_{\textrm{G}}(T)$ is the saturated vapor density of the helium gas,  $A$ is the cross-section of helium atom, and the matrix element $B_{mn}$ is given by

\begin{equation}
\frac{1}{B_{mn}}=\int\limits_{0}^{\infty} \Bigl \{ \psi_m (z) \psi_n (z) \Bigr \}^2 dz.
\label{eq:B}
\end{equation}

\noindent We note that for $m=n$, $\nu_{mn}$ coincides with the electron momentum relaxation rate for the corresponding intra-subband scattering. In addition, the following relation is valid for any arbitrary indexes $m$ and $n$

\begin{equation}
\nu_{nm}=\nu_{mn} \textrm{exp} \biggr ( -\frac{\epsilon_{m}-\epsilon_{n}}{T_{\textrm{e}}} \biggl ),
\label{eq:relation}
\end{equation}

\noindent where $\epsilon_{n}$ is given (in units of temperature) by (\ref{eq:wave}), and $T_{\textrm{e}}$ is the electron temperature.

In order to calculate the electron temperature and to describe the electron transport properties, we need to know the expressions for the momentum and the energy relaxation rates of the SSE, which are distributed over many subbands. In the case when the distribution function is given by the Boltzmann distribution, such expressions were obtained by Saitoh and Aoki.~\cite{Saitoh_hot} However, in the case of the MW absorption, when the transitions between two lowest subbands are induced by the MW radiation, it is necessary to generalize their equations for arbitrary distribution function. In this case, the momentum relaxation rate $\nu$ is given by~\cite{Monarkha}

\begin{equation}
\nu=\sum\limits_{m,n} \rho_{m} \nu_{mn} \textrm{exp} \biggr ( -\frac{\mid \epsilon_{m}-\epsilon_{n}\mid +\epsilon_{n}-\epsilon_{m}}{2T_{\textrm{e}}} \biggl ) \biggr[ 1+\frac{\mid \epsilon_{m}-\epsilon_{n}\mid +\epsilon_{n}-\epsilon_{m}}{2T_{\textrm{e}}} \biggl ],
\label{eq:rate}
\end{equation}

\noindent where $\rho_n$ is the fractional occupancy of the subband of index $n$. The corresponding expression for the energy relaxation rate $\tilde{\nu}$ is given by

\begin{equation}
\tilde{\nu}=\frac {m_{\textrm{e}}} {M} \sum\limits_{m,n} \rho_{n} \nu_{mn}
\textrm{exp} \biggr ( -\frac{\mid \epsilon_{m}-\epsilon_{n} \mid +\epsilon_{m}-\epsilon_{n}}{2T_{\textrm{e}}} \biggl ) \biggr [ 2+\frac{\mid \epsilon_{m}-\epsilon_{n} \mid}{T_{\textrm{e}}}+\frac{\hbar^2 B_{mn}}{2m C_{mn} T_{\textrm{e}}} \biggl ],
\label{eq:loss}
\end{equation}

\noindent  where $M$ is the mass of the helium atom, $B_{mn}$ is given by (\ref{eq:B}), and $C_{mn}$ is given by

\begin{equation}
\frac{1}{C_{mn}}=\int\limits_{0}^{\infty} \biggl \{ \frac{\textrm{d}}{\textrm{d}z}[\psi_m (z) \psi_n (z)] \biggr \}^2 dz.
\label{eq:C}
\end{equation}

\noindent Note that both rates reduce to the corresponding rates given in Ref.~7 in the case of Boltzmann distribution $\rho_n\propto \textrm{exp}(-\epsilon_n/T_{\textrm{e}})$.

\section{Resonance-induced resistivity}
\subsection{Experiment}

The experiment was done in the same experimental cell that was used for our direct MW absorption measurements 
\cite{Isshiki}. The cell is attached to the mixing chamber of the dilution refrigerator and the temperature of the mixing 
chamber is measured with a calibrated germanium thermometer. The cell temperature is assumed to be the same with that of 
the mixing chamber. 

A layer of the SSE is trapped on a vapor-liquid interface placed approximately midway between two parallel metal 
plates separated by 3~mm. By measuring the capacitance between the plates, the liquid level can be adjusted with the accuracy of about 0.05~mm.~\cite{Isshiki} Electrons are pressed toward the surface by an electric field created between the plates by applying a positive voltage $V_\textrm{B}$ to the bottom plate. By varying the potential of the bottom plate, the energy difference between the first excited and the ground states can be adjusted to match the frequency of the microwave field. The MW radiation of fixed frequency 130 GHz from a MW source (Gunn oscillator) is coupled into the cell by a dielectric waveguide of rectangular cross section. The microwave power dissipated in the cell can be estimated from the cooling power of the refrigerator and was usually $\sim$10~$\%$ of the input power $P$, which is measured at the output of the MW source.
  
The variation of the SSE resistivity is detected by a capacitive-coupling method~\cite{Shirahama}. For this purpose the top plate contains a concentric-ring copper electrode pair, which is known as a Corbino disk. The diameters of inner and outer electrodes are 14 and 20~mm respectively, and the gap between them is less than 0.2~mm. To charge the surface, a positive voltage $V_\textrm{B}$ in a range from 3 to 25~V is applied to the bottom electrode. The electrons, which are produced at 0.65~K by thermionic emission from a tungsten filament located approximately 1.5~mm above the surface, accumulate on the surface until the charge layer completely screens the electric field above the surface. This condition determines the charge density $n_e=\varepsilon V_\textrm{B} /4\pi e d$, where $d$ is the depth of the liquid. An ac voltage of 1~MHz and 3$-$5~mV$_\textrm{RMS}$ is applied to the inner electrode. This produces an in-plane electric field $E_{\parallel}$, which drives electrons parallel to the liquid surface. The current induced in the SSE is picked up at the outer electrode, and the amplified signal is synchronously demodulated at 1~MHz by means of a lock-in amplifier. From the value of this current the magnitude of $E_{\parallel}$ was estimated to be less than $10^{-3}$~V/cm, which should produce no significant heating of the SSE in the temperature range of this experiment~\cite{Saitoh_warm}. The resistivity of the SSE is found from the values of the phase shift of the output signal with respect to the driving signal using fitting procedure described in detail elsewhere~\cite{Shirahama}.

To observe the resonance-induced change of the resistivity, the voltage $V_\textrm{B}$ is varied to tune the SSE in resonance with MW radiation, and the in-phase and quadrature components of the output Corbino signal are recorded. The input MW power $P$ is varied in a range from about 10~$\mu$W, at which the Corbino signal can be distinguished from the noise (we would like to notice that our direct absorption setup described in Ref.~4 allows much more sensitive detection of MW absorption than the method described in this section), to about 1500~$\mu$W, which is the maximum output power of our MW source.

\subsection{Results}

\begin{figure}[t]
\begin{center}
\includegraphics[width=12cm]{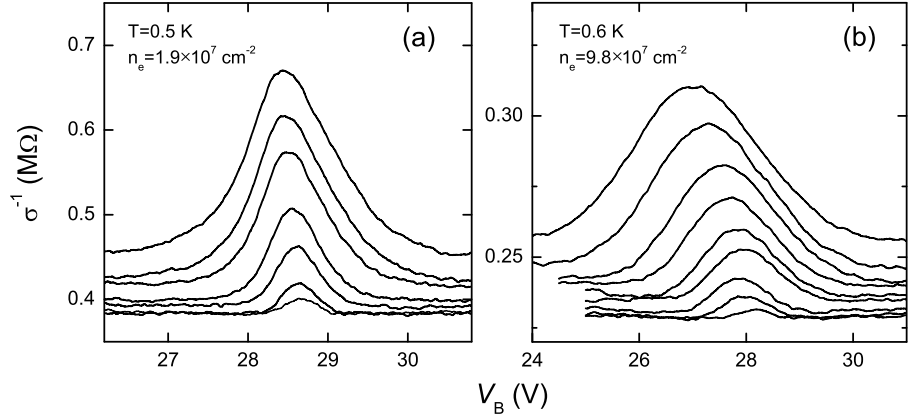}
\end{center}
\caption{\label{fig:exp2} Variation of the  resistivity $\sigma^{-1}$ of the SSE taken at (a): T=0.5 K, n=1.9$\times10^7$~cm$^{-2}$ and (b): T=0.6 K, n=9.8$\times10^7$~cm$^{-2}$. The resistivity curves in (a) correspond (from bottom to top) to 20, 40, 95, 190, 400, 670, and 1085~$\mu$W of the input MW power, while the curves in (b) correspond (from bottom to top) to 50, 95, 120, 190, 300, 400, 670, 1080, and 1500~$\mu$W.}
\end{figure}

The resistivity measurements were done at temperatures from 0.65~K down to 0.45~K, i.e. in the vapor-atom scattering regime. The experimental procedure described in previous section allowed us to plot the variation of resistivity $\sigma^{-1}$ with $V_\textrm{B}$ at different values of input MW power. Examples of such plots for two different temperatures are shown in Fig.~\ref{fig:exp2}. The data at 0.5 and 0.6~K were taken at electron densities of $1.9\times 10^7$ and $9.6\times 10^7$~cm$^{-2}$ respectively. The resonant change of the resistivity is due to the resonant excitation of the SSE from the ground to the first excited subband, with peak value at $V_\textrm{B}\approx$ 28.6~V at small MW power (there is a shift of about 0.4~V in the resonance voltage at high electron densities, which is attributed to the effect of the image charges induced in metal plates~\cite{shift}). The relation between the pressing electric field $E_{\perp}$ and the transition frequency $\omega_{12}/2\pi$ can be calculated from the numerical solution of (\ref{eq:wave}) and at the typical values of $E_{\perp}$ used in our experiment can be approximated by $\omega_{12}/2\pi=82.2+0.5E_{\perp}$, where frequency is in GHz and electric field is in V/cm. This estimate is in good agreement with value of $V_\textrm{B}$ at the resonance observed in our experiment.

The excitation of the inter-subband transition causes $\sigma^{-1}$ to increase, as was observed at all temperatures between 0.45 and 0.65~K and at all values of $P$. The increase of $\sigma^{-1}$ far from the resonance with increasing $P$, which can be seen in Fig.~\ref{fig:exp2} as the vertical shift of signals toward higher values of $\sigma^{-1}$, is attributed to the heating of the experimental cell by MW radiation. The increase of the cell temperature can be estimated from the measured temperature dependence of the SSE conductivity and is $\sim10^{-2}$~K at the highest value of $P$. 

It is important to mention that the MW radiation entering our experimental cell is not vertically polarized, therefore there should be a component of MW electric field which drives electrons parallel to the surface. The heating of the SSE due to this field can be estimated as follows. The average power absorbed by the SSE from the driving field of amplitude $E_{\parallel}$ is about $\sigma E_{\parallel}^2$. In the case of the low-frequency field, the heating is expected to become noticeable at $E_{\parallel}\gtrsim 10^{-2}$~V/cm at $T\approx 0.5$~K.~\cite{Saitoh_warm} However, for such high-frequency field as $\omega\tau\gg 1$, where $\tau$ is the electron scattering time, the conductivity of electrons is reduced by a factor of $(\omega\tau)^2$ comparing with the dc value. For the typical scattering time of the order of $10^{-9}$~s at $T\approx 0.5$~K, this factor is more than $10^5$ for MW frequency of 130~GHz. Therefore, in the temperature range of the experiment, the heating of SSE due to parallel field is estimated to be negligible even at the highest values of $P$. 

At high MW power, the resonance line shape becomes slightly asymmetric with the left hand side being slightly steeper than the right hand side. Also, the position of the peak amplitude of the resistivity curves shifts toward the lower values of $V_\textrm{B}$ with increasing input power. The absolute value of the frequency shift was found to increase with the electron density. For example, for two sets of data shown on Fig.~\ref{fig:exp2}, which correspond to density values of $1.9\times 10^7$ and $9.8\times 10^7$~cm$^{-2}$, the shifts are estimated to be about 0.4 and 1.9~GHz respectively.  Variation of the liquid level due to thermal expansion of the liquid could be one of the possible explanations of this shift. MW radiation heats up the cell and variation of the liquid level can cause the change of the magnitude of pressing field due to image charges induced in the metal plates. However, using the values of the thermal expansion coefficient of $^3$He~\cite{Roach}, the effect is estimated to be several orders of magnitude smaller. Another exciting possibility is the shift of the resonance frequency due to Coulomb interaction between electrons~\cite{Lea}. For the SSE, the average distance to the surface increases with the subband index, and the energy levels of an electron shift when the neighbour electron is excited, because the distance between electrons also changes. As a result, the transition frequency of the ground state electron increases with increasing population of the excited levels. The shift is larger near the resonance than far from it, thus introducing the asymmetric into the line shape, which is indeed observed in the experiment. The absolute value of the frequency shift is expected to increase with electron density, which is in agreement with above observation. The value of this shift can be easily estimated for two isolated electrons. Using formula similar to the one in Ref.~21, for two electrons above $^3$He in zero field, and separated by 1~$\mu$m, we have frequency shift of about 0.9 GHz, which is within an order of magnitude agreement with our observation. For more realistic estimation, the interaction between many electrons has to be considered, and long-range Coulomb interaction between distant electrons might need to be taken into account. This is a rather complicated problem, therefore we postpone any further discussion of this effect until future publications.

\subsection{Discussion}
\subsubsection{Hot electron theory}

Let us suppose that as a result of the MW excitation a fraction $\rho_2$ of the SSE occupy the first excited state. First, let us assume that the SSE are in thermal equilibrium with the vapor, i.e. have the same temperature $T$ as the vapor and liquid. Since $T$ is much less than the energy difference between the ground and the excited levels, we can ignore the thermal population of the higher index subbands, i.e. consider only scattering of electrons within two lowest subbands. From (\ref{eq:rate}), the momentum relaxation rate $\nu$ in this case is given by

\begin{equation}
\nu=\rho_1 \nu_{11}+\rho_2 \nu_{22}+\rho_2 \nu_{21}+\rho_1 \nu_{12} \biggr ( 1+\frac{\epsilon_2-\epsilon_1}{T_{\textrm{e}}} \biggl ),
\label{eq:2level}
\end{equation}

\noindent where $\rho_1=1-\rho_2$, and $\nu_{mn}$ are given by (\ref{eq:relation}). Since $\nu_{12}\ll \nu_{21}$, the last term in (\ref{eq:2level}) can be safely neglected. The numerical estimate shows that the relaxation rate monotonically decreases with increasing $\rho_2$. For example, at $T=0.5$ K and the pressing field $E_{\perp }\cong 95$~V/cm, at which $\nu_{11}$, $\nu_{22}$ and $\nu_{21}$ are about $2.0\times 10^9$, $1.1\times 10^9$ and $0.4\times 10^9$ s$^{-1}$ respectively, the relaxation rate decreases by about $3\%$ and $12\%$ at the populations $\rho_2$ of 0.1 and 0.5 respectively. Thus, if we neglect the heating of the SSE, the resistivity is expected to drop at the resonance. This is opposite to the behaviour observed in the experiment. 

However, the assumption that the SSE are in thermal equilibrium with the vapor can not be justified if electrons absorb large energy from an external source. On the one hand, due to very large (order of $10^{11}$~s$^{-1}$) electron-electron collision rate the SSE can rapidly re-distribute the energy among themselves, thus quickly attaining an equilibrium with effective electron temperature $T_{\textrm{e}}$. On the other hand, due to very weak coupling of the SSE to the environment, their temperature can be much higher than the temperature $T$ of the vapor and liquid. When electrons become hot and thermally populate many subbands, the inter-subband scattering among the occupied levels contribute significantly to the scattering rate. In other words, the scattering rate of the electron increases with the density of states to which it can be scattered, and the latter increases rapidly as more and more subbands become populated. As a result, in the gas atom scattering regime the momentum relaxation rate, given by equation (\ref{eq:rate}), is almost monotonically increasing function of the electron temperature $T_{\textrm{e}}$. Thus, the significant rise of the electron resistivity observed in our experiment can be explained if we assume that the SSE become hot.The heating of the SSE can be caused by the absorbed MW power and is due to very slow energy relaxation rate. The life time of the excited electron is mainly limited by the elastic collisions with helium vapor atoms. The energy exchange between the electron and the helium atom during one collision is negligible due to large difference in masses. As a result, the electron returns to the ground state carrying large kinetic energy, which is quickly distributed between all other electrons in the ground subband. The cooling of electrons is a slow process and the temperature of electrons can rise well above that of the helium bath.

In order to find $T_{\textrm{e}}$ as a function of the MW power in the cell, we write the energy balance equation by equating the average energy absorbed by the electron from the MW field in the cell (per unit time) to the average energy lost in the collisions with helium atoms, i. e.

\begin{equation}
\frac{0.5\hbar \omega \gamma \Omega^2}{(\omega_{12} -\omega)^2+\gamma^2} ( \rho_1 -\rho_2 ) = ( T_{\textrm{e}}-T )\tilde{\nu}.
\label{eq:bal}   
\end{equation}

\noindent where $\omega_{12}$ is the transition frequency, which is the function of the pressing field $E_{\perp}$, $\omega$ is the MW frequency, $\gamma$ is the temperature dependent transition linewidth, $\Omega$ is the Rabi frequency defined below, and the energy relaxation rate $\tilde{\nu}$ is given by equation (\ref{eq:loss}). In this equation, we neglect the frequency shift observed in the experimental data and discussed at the end of Sec.~3.2.

In general case, the linewidth $\gamma$ depends not only on the ambient temperature $T$, which determines the vapor density $N_G$, but also on the electron temperature $T_{\textrm{e}}$. The expression for $\gamma$, which takes into account the scattering of the SSE from two lowest subbands into higher index ($n>2$) subbands, was found by Ando.\cite{Ando} Using the notations adopted earlier, it can be written as

\begin{equation}
\gamma=\frac{1}{2}\bigr( \nu_{11}+\nu_{22}-\nu_{21}+\nu_{12}+\sum\limits_{n>2}\nu_{1n}+\sum\limits_{n>2}\nu_{2n}  \bigl).
\label{eq:gamma}
\end{equation}

\begin{figure}[t]
\centering
\includegraphics[width=11cm]{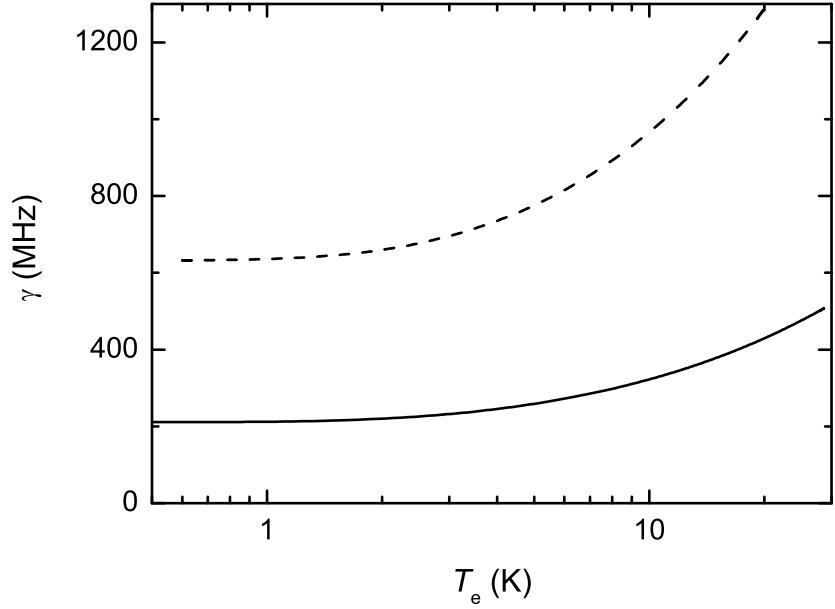}
\caption{\label{fig:gamma} Variation of the linewidth $\gamma$ with $T_{\textrm{e}}$ calculated for $T$=0.5~K (solid line) and $T$=0.6~K (dashed line).}
\end{figure}

\noindent In Fig.~\ref{fig:gamma} we show the plots of $\gamma$ as a function of $T_{\textrm{e}}$ calculated for $T$=0.5~K (solid line) and $T$=0.6~K (dashed line). At $T$=0.5~K we find that $\gamma$ increases from about 210~MHz at $T_{\textrm{e}}$=0.5~K to about 520~MHz at $T_{\textrm{e}}$=28.7~K.

\begin{figure}[t]
\centering
\includegraphics[width=11cm]{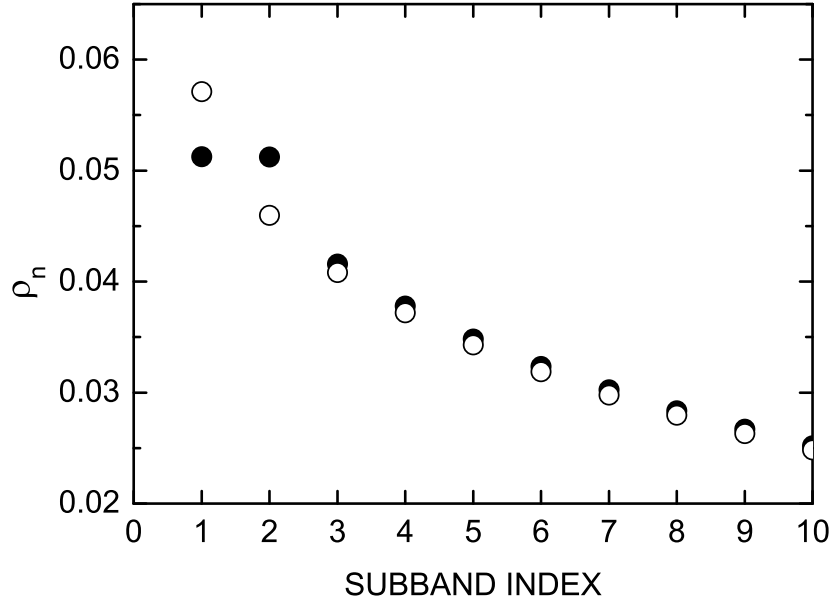}
\caption{\label{fig:popul} Comparison between subband occupancies $\rho_n$ given by Boltzmann distribution (open circles)  and that found by numerical solution of rate equations (\ref{eq:popul}) (solid circles) at $T=0.5$~K, $T_{\textrm{e}}=28.7$~K and $\Omega=10$~GHz.}
\end{figure}

The fractional occupancies $\rho_{n}$ are determined by the balance between the rate of the MW excitation and the scattering rates $\nu_{mn}$. At the stationary state, i.e. $\partial \rho_{n}/\partial t$=0, they can be found from the following equations

\begin{subequations}
\begin{eqnarray}
\frac{0.5\gamma \Omega^2}{(\omega_{12} -\omega)^2+\gamma^2} ( \rho_2 -\rho_1 )-\sum\limits_{n\neq 1}\nu_{1n}\rho_1+\sum\limits_{n\neq 1}\nu_{n1}\rho_n=0,\label{eq:popula}
\\
\frac{0.5\gamma \Omega^2}{(\omega_{12} -\omega)^2+\gamma^2} ( \rho_1 -\rho_2 )-\sum\limits_{n\neq 2}\nu_{2n}\rho_2+\sum\limits_{n\neq 2}\nu_{n2}\rho_n=0,\label{eq:populb}
\\
-\sum\limits_{n\neq m}\nu_{mn}\rho_m+\sum\limits_{n\neq m}\nu_{nm}\rho_n=0,\label{eq:populc}
\end{eqnarray}
\label{eq:popul}
\end{subequations}

\noindent where equation (\ref{eq:populc}) is valid for all $m\geq 3$. From the above equations, the occupancies $\rho_{n}$ can be calculated as a function of $T_{\textrm{e}}$ and the MW power. Like in Ref.~7, in our calculations we assume that the electron temperature $T_{\textrm{e}}$ is the same for all subbands of indexes $n$ (effective temperature approximation). The dependence on the MW power can be expressed in terms of the Rabi frequency. The latter is defined as $\Omega=ez_{12}E_{\textrm{RF}}/\hbar$, where $z_{12}=\langle \psi_1 |z| \psi_2 \rangle$  and $E_{\textrm{RF}}$ is the amplitude of the MW field in the cell, and is proportional to the root square of the MW power in the cell. In order to obtain $\rho_{n}$, equations (\ref{eq:popul}) were solved numerically for large number of $n\leq n_{\textrm{max}}$. We have checked the convergence of the result by taking $n_{\textrm{max}}$=200 and 400, and the result for $n_{\textrm{max}}$=400 is presented here. The thermal population of subbands with indexes $n>n_{\textrm{max}}$ was neglected. To determine the energies $\epsilon_n$ and the wave functions $\psi_n(z)$ of the electron in the non-zero pressing filed, we solved the corresponding Schrodinger equation numerically for quantum numbers $n\leq$10. In this equation, the surface barrier potential due to liquid was taken to be infinite. For $n>$10 the image potential was neglected and the solutions given by Airy functions were used. The examination of the obtained solutions of equations (\ref{eq:popul}) showed that at $\Omega \gtrsim 10$~MHz, the fractional occupancies $\rho_{n}$ for $n \geq 3$ deviate very little (less then $4\%$) from those given by the Boltzmann statistics, i.e.

\begin{equation}
\rho_{n}=\frac{1}{Z}\textrm{exp} \biggr( -\frac{\epsilon_{n}}{T_{\textrm{e}}} \biggl),
\label{eq:Boltz}
\end{equation}

\noindent where $Z$ is the partition function. At the same time, the populations of the two lowest levels depend on the MW power and reach the saturation, i.e. $\rho_2\rightarrow \rho_1$, at high values of $\Omega$. As an example, in Fig.~\ref{fig:popul} we show the comparison between $\rho_{n}$ calculated from (\ref{eq:popul}) and those given by (\ref{eq:Boltz}) at some values of $T_{\textrm{e}}$ and $\Omega$.

\begin{figure}[t]
\centering
\includegraphics[width=11cm]{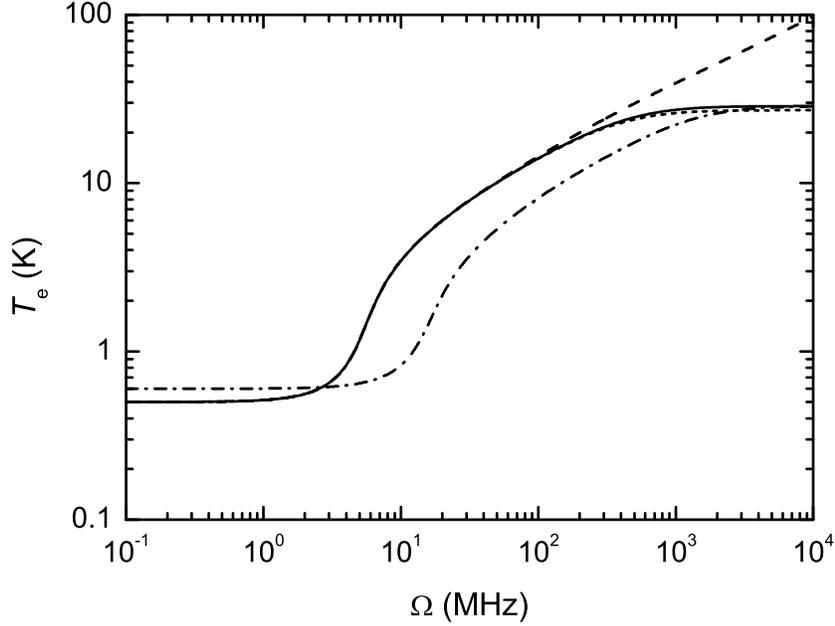}
\caption{\label{fig:temp} Electron temperature versus Rabi frequency calculated for electrons above $^3$He in the pressing field $E_{\perp }$=95.33~V/cm for $T$=0.5~K (solid line) and $T$=0.6~K (dash-dotted line). Short-dashed line and dashed line are the calculations of $T_{\textrm{e}}$ which assume Boltzmann distribution for $n\geq 3$ and for $n\geq 1$ ("pure" Boltzmann distribution) respectively. Both lines are calculated for $T$=0.5~K.}
\end{figure}

The electron temperature $T_{\textrm{e}}$ can be calculated as a function of the Rabi frequency $\Omega$ from equations (\ref{eq:bal}) and (\ref{eq:loss}) taking the sum over $n\leq n_{\textrm{max}}$. The plot of $T_{\textrm{e}}$ versus $\Omega$ calculated for $T$=0.5~K (solid line) is shown in Fig.~\ref{fig:temp}. At small values of $\Omega$, the heating of the SSE is not significant, and the electron temperature stays close to that of the liquid. At $\Omega \gtrsim 2$~MHz, the power absorbed by the SSE increases, the cooling rate is too small to keep electrons in the thermal equilibrium with the vapor, and $T_\textrm{e}$ rises quickly. At $T_\textrm{e} \gtrsim 2$~K, the thermal population of the higher levels become appreciable, the cooling rate increases, and the electron temperature rises slower. At very high values of $\Omega$, $\rho_2$ approaches $\rho_1$ as was shown above, the power absorption saturates, $T_{\textrm{e}}$ reaches maximum value of about 28.7~K and becomes almost power independent. The plot calculated for $T$=0.6~K (dash-dotted line) is also shown in Fig.~\ref{fig:temp}. At this temperature the vapor density $N_{\textrm{G}}$, which determines the scattering rates, is higher than at $T$=0.5~K, therefore the cooling of the SSE is more efficient. As a result, $T_{\textrm{e}}$ starts to rise at higher MW power ($\Omega \gtrsim 10$~MHz), and stays less than $T_{\textrm{e}}$ calculated for $T=0.5$~K. However, at saturation it reaches almost the same maximum value (about 28.6~K) as for the $T=0.5$~K. This result is not surprising and can be expected from equations (\ref{eq:populb}) and (\ref{eq:bal}) if we take $\omega_{12}=\omega$. At saturation, the fractional occupancies $\rho_n$ reaches some constant values for all $n$. In this case, it is easily seen that $(\rho_1-\rho_2)$ is proportional to the temperature dependent vapor density ${N_G}^2(T)$. Since both the linewidth $\gamma$ and the energy relaxation rate $\tilde{\nu}$ are proportional to $N_G$, the dependence on ambient temperature $T$ cancels out from equation (\ref{eq:bal}) at $T_{\textrm{e}}\gg T$. Thus, it gives the same $T_{\textrm{e}}$ regardless of the value of $T$.   
 
\begin{figure}[t]
\centering
\includegraphics[width=11cm]{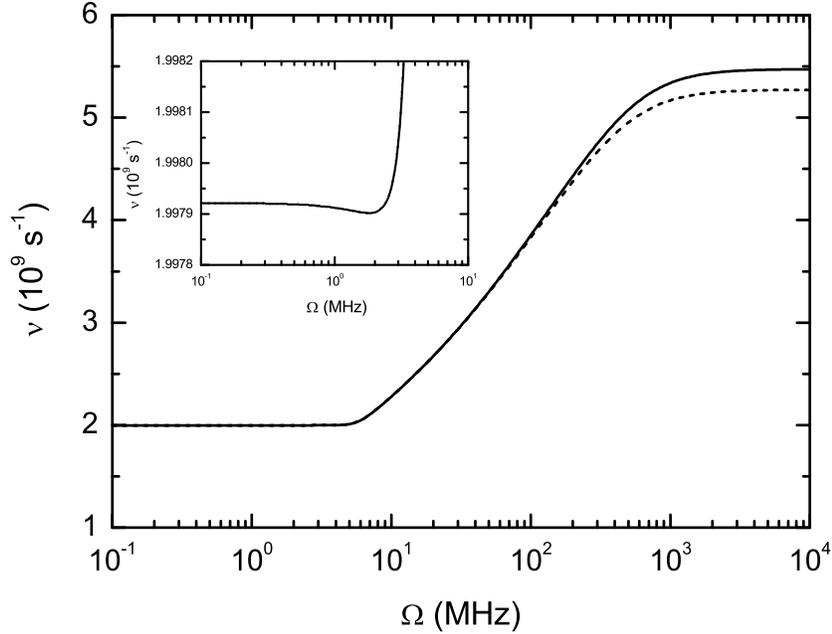}
\caption{\label{fig:rate} The momentum relaxation rate (solid line) due to scattering by helium vapor atoms versus Rubi frequency calculated for the SSE above $^3$He in the pressing field $E_{\perp }$=95.33~V/cm. Sort-dashed line is the calculation of $\nu$ which assumes Boltzmann distribution for $n\geq 3$. Both lines calculated for $T$=0.5~K.}
\end{figure}  

In order to see the effect of the heating on the electron resistivity, we have calculated the momentum relaxation rate $\nu$ of the SSE as a function of $T_{\textrm{e}}$ and $\Omega$ using equation (\ref{eq:rate}). The plot of $\nu$ versus $\Omega$ calculated at $T$=0.5~K is shown in Fig.~\ref{fig:rate}. At small $\Omega$, that is at small MW power, when the increase of $T_e$ is negligible, the relaxation rate slightly decreases, as shown in the inset of Fig.~\ref{fig:rate}. This result is consistent with the condition considered under equation (\ref{eq:2level}), that is $\nu_{12}\ll \nu_{21}$. At $\Omega \gtrsim 5$~MHz, electrons start thermally populating the higher excited states, the inter-subband scattering increases and the relaxation rate quickly rises. At large values of $\Omega$, when $T_{\textrm{e}}$ becomes power independent (see Fig.~\ref{fig:temp}), the relaxation rate reaches constant value of about $5.5\times 10^{9}$~s$^{-1}$. Therefore, if we neglect the small decrease (less then 0.001$\%$) at $\Omega \lesssim 2$~MHz, the relaxation rate increases monotonically with the MW power, which qualitatively agrees with our experimental result. At $T$=0.6~K (plot is not shown) the relaxation rate is about $6\times 10^{9}$~s$^{-1}$ for cold electrons and reaches about $14\times 10^{9}$~s$^{-1}$ for hot electrons at saturation.

As was shown earlier, at high enough values $T_{\textrm{e}}$ the populations of the levels with $n\geq 3$ are close to those given by the Boltzmann statistics. Therefore, to simplify the calculations and to save the computation time one can use (\ref{eq:Boltz}) to calculate $\rho_n$ for all $n\geq 3$, while determine the population of the two lowest levels from one of the rate equations, e.g. (\ref{eq:popula}), and the condition that $\sum\limits_{n}^{\infty} \rho_n=1$. In this case, the sums appearing in equations (\ref{eq:bal}) and (\ref{eq:rate}) can be taken over all values of index $n$ using the asymptotic formulas given in Appendix of Ref.~7. The results of these calculations for $T$=0.5~K are shown as short-dashed lines in Figs.~\ref{fig:temp} and \ref{fig:rate}. In this case $T_{\textrm{e}}$ and $\nu$ reach maximum values of about 27~K and $5.3\times 10^{9}$~s$^{-1}$ respectively.

It is also instructive to calculate $T_{\textrm{e}}$ assuming "pure" Boltzmann distribution of the SSE over the subbands, i.e. assuming that $\rho_n$ are given by (\ref{eq:Boltz}) for all indexes $n\geq 1$. This is the situation considered by Saitoh and Aoki\cite{Saitoh_hot} and, most recently, by Ryvkine \textit{et al}.\cite{Ryvkine} For the pure Boltzmann distribution, the plot $T_{\textrm{e}}$ versus $\Omega$ is also shown in Fig.~\ref{fig:temp} (dashed line). As should be expected for Boltzmann distribution, for large $\Omega$ the occupancy of the first excited state $\rho_2$ approach the occupancy $\rho_1$ of the ground state much slower than the corresponding solutions of the rate equations (\ref{eq:rate}). As a result, the MW absorption does not saturate, and $T_{\textrm{e}}$ monotonically increases with the power, which is similar to the result obtained for a different heating mechanism in Ref.~7.

\begin{figure}[t]
\centering
\includegraphics[width=11cm]{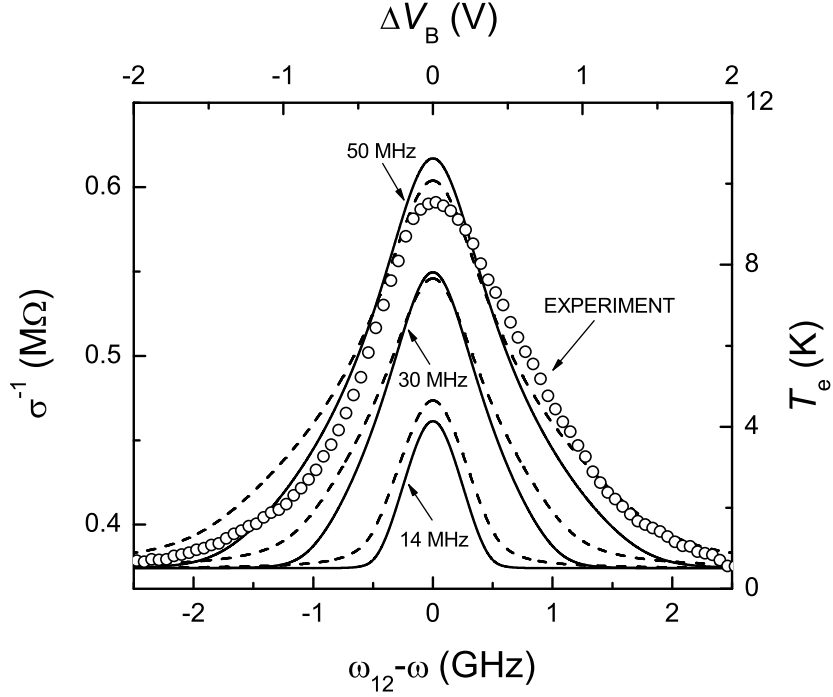}
\caption{\label{fig:res} Variation of $\sigma^{-1}$ (solid line) and $T_{\textrm{e}}$ (dashed line) with the transition frequency $\omega_{12}$ calculated for $T$=0.5~K, $n_\textrm{e}=1.9 \times 10^{7}$ cm$^{-2}$ and for three different values of $\Omega$: 14, 30 and 50~MHz. Experimental data taken at $T=0.5$~K and $P=1085$~$\mu$W are potted by
 symbols as a function of voltage $V_{\textrm{B}}$.}
\end{figure}   

\subsubsection{Comparison with experiment}     

In order to make a quantitative comparison between the experimental curves shown in Fig.~\ref{fig:exp2} and the predictions of our model, we have calculated the variation of $\sigma^{-1}$ with the transition frequency $\omega_{12}$ at different values of the Rabi frequency, and compared the calculated curves with the experimental ones. First, we calculate the variation of $T_{\textrm{e}}$ with the frequency $\omega_{12}$ at a fixed values of $\Omega$ and $\omega$ from equations (\ref{eq:bal}) and (\ref{eq:loss}). Once the electron temperature is known, the corresponding variation of resistivity $\sigma^{-1}(\omega_{12})$ and change of resistivity $\Delta \sigma^{-1}(\omega_{12})$ can be calculated using (\ref{eq:rate}) and the relation $\sigma^{-1}=m_{\textrm{e}} \nu/(n_{\textrm{e}}e^2)$. As an example, in Fig.~\ref{fig:res} we show the variation of  $\sigma^{-1}$ and $T_{\textrm{e}}$ with $\omega_{12}-\omega$ calculated for $T$=0.5~K, $n_\textrm{e}=1.9 \times 10^{7}$ cm$^{-2}$ and several values of $\Omega=$14, 30 and 50~MHz. As expected, $\sigma^{-1}$ monotonically increases as $\omega_{12}$ approaches MW frequency $\omega$, reaching the maximum value at the resonance. All curves calculated for $\Omega \gtrsim 5$~MHz and $T=0.5$~K show similar behaviour. For comparison, the experimental data for variation of $\sigma^{-1}$ with $\Delta V_{\textrm{B}}=V_{\textrm{B}}-{V_{\textrm{B}}}^0$, where ${V_{\textrm{B}}}^0$ is the bottom electrod voltage at which $\sigma^{-1}$ reaches maximum, is also plotted in Fig.~\ref{fig:res}.
 
\begin{figure}[t]
\centering
\includegraphics[width=11cm]{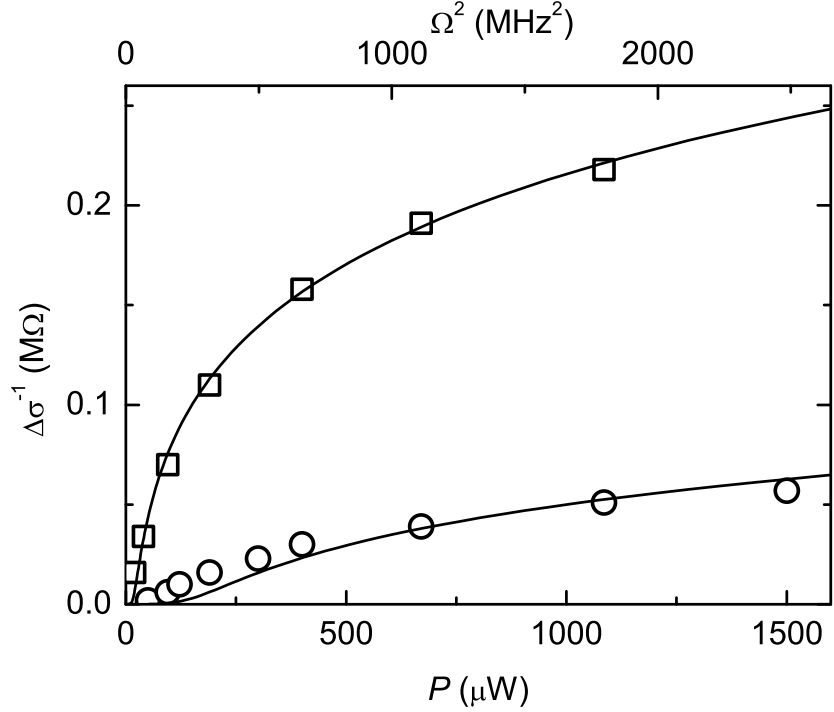}
\caption{\label{fig:comp} Peak amplitude of resistivity curves plotted in Fig.~\ref{fig:exp2}(a,b) vs input power $P$ indicated on bottom axis (squares for $T$=0.5~K and circles for $T$=0.6~K), and theoretical calculations of $\Delta \sigma^{-1}$ vs $\Omega^2$ indicated on the top axis (solid lines).}
\end{figure}
 
The comparison between the peak amplitudes of the experimental and the theoretical resistivity curves calculated for $T$=0.5 and 0.6~K is given in Fig.~\ref{fig:comp}. The squares and circles represent the peak amplitude of the curves shown in Figs.~\ref{fig:exp2}(a) and \ref{fig:exp2}(b) respectively and plotted as a function of the input MW power $P$, while the calculated lines are plotted as a function of $\Omega^2$. Although the magnitude of the MW power inside the cell is not measured, we find it reasonable to assume that it is proportional to $P$. Because the proportionality coefficient between $\Omega$ and $\sqrt{P}$ is not known, the range of the horizontal axis for the theoretical line is adjusted to give the best fit between the experimental and the calculated plots for $T$=0.5~K. The vertical axis is the same for both experimental and theoretical plots. From this procedure the proportionality coefficient between $\Omega$ and $\sqrt{P}$ is found to be about 1.26$\times 10^3$ MHz/$\sqrt{\textrm{W}}$. Then, $\Omega$ is estimated to be about 40~MHz at the highest input power for the data shown in Fig.~\ref{fig:exp2}(a). As seen in Fig.~\ref{fig:comp}, this estimate also give a reasonable agreement between the experimentally observed and theoretically calculated resistivity change at $T$=0.6~K. From this result we conclude that at the input power of about 1000 $\mu$W SSE are heated up to about 9~K and to about 5~K for T=0.5 and 0.6~K respectively.

\section{Absorption line broadening}
\subsection{Introduction}

In addition to the electrical resistivity, MW absorption-induced heating is also expected to effect the line shape of the absorption signal. The power absorption $P_{\textrm{A}}$ depends on the fractional occupancies $\rho_1$ and $\rho_2$ of two lowest subbands and in general case is given by

\begin{equation}
P_{\textrm{A}}=\frac{0.5\hbar \omega \gamma \Omega^2}{(\omega_{12} -\omega )^2+\gamma^2} ( \rho_1 -\rho_2 ).
\label{eq:abs}
\end{equation}

\noindent For cold electrons ($T_e\approx T$), thermal population of higher excited subbands can be neglected. Taking into account the occupation of only two lowest levels, i.e. $\rho_1+\rho_2=1$, it is straightforward to show from (\ref{eq:rate}) that equation (\ref{eq:abs}) reduces to~\cite{Collin} 

\begin{equation}
P_{\textrm{A}}=\frac{0.5\hbar \omega \gamma \Omega^2}{(\omega_{12} -\omega )^2+\gamma^2+\gamma\tau\Omega^2},
\label{eq:absCOLD}
\end{equation}

\begin{figure}[t]
\centering
\includegraphics[width=11cm]{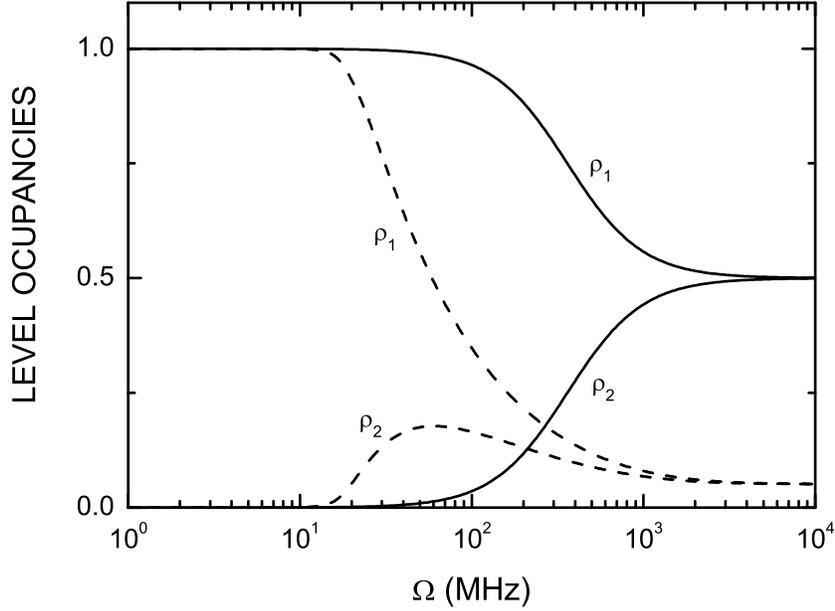}
\caption{\label{fig:rho} Fractional occupancies $\rho_1$ and $\rho_2$ verses $\Omega$ calculated at $T=0.6$~K for cold electrons (solid lines) and for hot electrons (dashed lines) using hot electron theory developed in Sec.~5.1.}
\end{figure}

\noindent Here $\tau=\nu_{21}^{-1}$ is the life time of an electron in the excited state. For cold electrons, $\gamma\tau$ is temperature independent and is about 3.08 for $E_{\textrm{z}}=95.33$ V/cm for SSE above $^3$He. According to (\ref{eq:absCOLD}), the absorption line shape is Lorenzian with half-width at half-maximum given by $(\gamma^2+\gamma\tau\Omega^2)^{1/2}$. For $\Omega\ll \gamma/(\gamma\tau)^{1/2}$, the absorption increases linearly with input MW power at all values of $\omega_{12}-\omega$, and the line shape is independent of input power $P$. In the opposite limit of large $\Omega$, the line shape becomes power broadened, and the half-width increases proportionally to $\sqrt{P}$. The reason for the power broadening is that at high MW excitation rates $\rho_2\rightarrow \rho_1$, as illustrated in Fig.~\ref{fig:rho} where we plot the fractional occupancies as a function of $\Omega$ (solid lines) calculated for $T=0.6$~K. In the limit of large $\Omega$, $(\rho_1-\rho_2)$ become proportional to $\Omega^{-2}$. As a result, the dependence on $\Omega$ cancels out from equation (\ref{eq:abs}) and $P_{\textbf{A}}$ saturates. Since the absorbed power increases slower near the resonance than far from the resonance, this leads to the broadening of the line.
 
\begin{figure}[t]
\centering
\includegraphics[width=11cm]{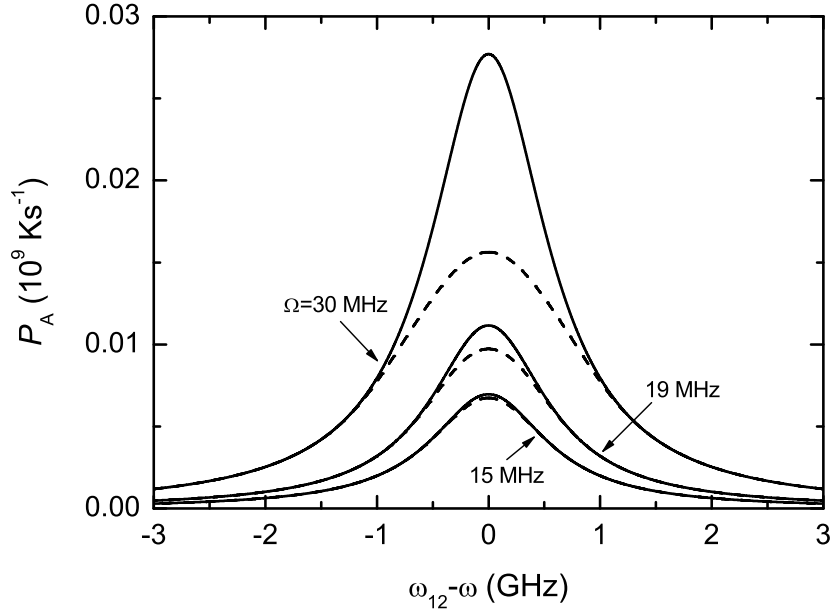}
\caption{\label{fig:bleach} Power absorption versus transition frequency $\omega_{12}$ calculated for $T$=0.6~K and for three different values of $\Omega$=15, 19 and 30~MHz. Solid lines are calculated ignoring heating and using (\ref{eq:absCOLD}), while dashed lines are for hot electrons using (\ref{eq:abs}).}
\end{figure}

However, the situation becomes quite different if electron heating is taken into account. In this case, $(\rho_1-\rho_2)$ starts decreasing with $\Omega$ due to the thermal excitations of the electrons from the ground state to the excited levels long before any noticeable MW excitation takes place. This is clearly seen in Fig.~\ref{fig:rho} where we plot the fractional occupancies for hot electrons (dashed lines) calculated using the hot electron theory developed in Sec.~5.1. As a result, the absorption line is expected to broaden long before the saturation condition is reached. This effect is illustrated in Fig.~\ref{fig:bleach} where we plot absorption lines calculated for $\Omega$=15, 19 and 30~MHz and $T$=0.6~K using the hot electron theory described in the previous section. The solid lines are calculated for cold electrons using (\ref{eq:absCOLD}), while dashed lines are for hot electrons using more general expression (\ref{eq:abs}). As seen in Fig.~\ref{fig:bleach}, at small values of $\Omega$ the heating is negligible, and the absorption is well described by (\ref{eq:absCOLD}). At higher values of $\Omega$, however, the absorption lines calculated for cold and hot electrons differ significantly in the vicinity of the resonance. The difference come from the fact that near the resonance, where the heating is most prominent, the reduction of $\rho_1$ and the enhancement of $\rho_2$ due to the thermal excitation leads to the decrease of $P_{\textrm{A}}(\omega_{12})$ in accordance with (\ref{eq:abs}). Far from the resonance, the heating is weak and $P_{\textrm{A}}(\omega_{12})$ is close to that given by (\ref{eq:absCOLD}). This leads to the additional broadening of the absorption line. We note that at $T$=0.6~K, the linewidth $\gamma$ calculated using Ando theory~\cite{Ando} is about 630~MHz (see Fig.~\ref{fig:gamma}), therefore the maximum value of $\Omega\cong 40$~MHz found in the experiment is too small to cause absorption saturation.

\begin{figure}[t]
\centering
\includegraphics[width=11cm]{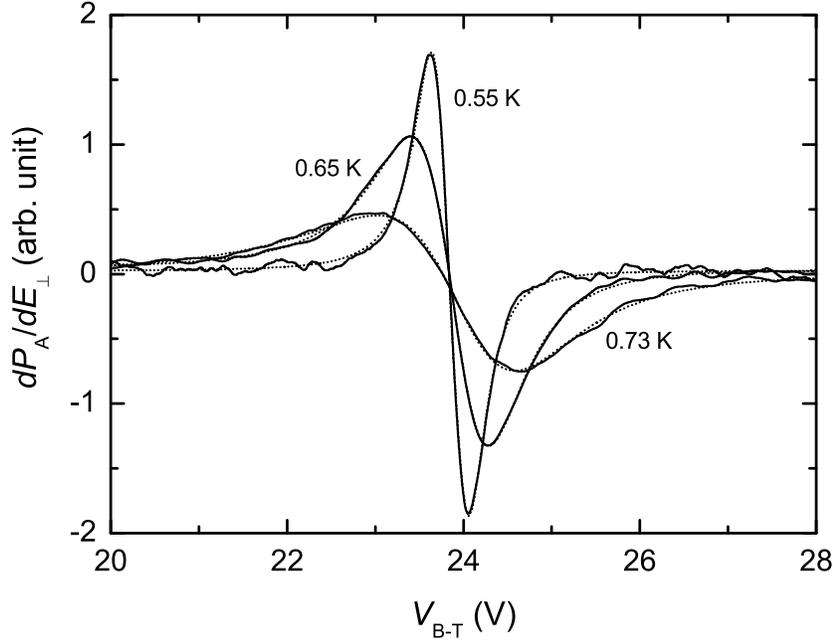}
\caption{\label{fig:deriv} Absorption derivative signals taken at $n$=10$^8$~cm$^{-2}$, $P$=5~$\mu$W and 0.55, 0.65 and 0.73~K (solid lines). The dotted lines are the fits obtained as described in the text.}
\end{figure}

\subsection{Experiment} 

In order to see the effect of heating on the absorption linewidth, we conducted an experiment in which we measured the absorption linewidth as a function of the input power. The experimental procedure for direct absorption measurements was similar to described in Ref.~4, although the experiment was done in a new experimental cell designed for magneto-resistivity measurements. In the new cell, SSE were accumulated on the liquid $^3$He surface placed half way between two parallel round conducting electrodes separated by 2.6~mm. As in the previous setup, the top plate consisted of Corbino disk used for magneto-resistivity measurements. In addition, the circular conducting guard rings were placed around each electrode. MW power at fixed frequency of 130~GHz was passed through the cell and was detected upon its exiting by means of the InSb bolometer mounted at the still of the dilution refrigerator. Before the measurements, the parallelism between helium surface and electrodes was adjusted to within about 1~mrad by tilting the cryostat and obtaining the Corbino signal which corresponded to the maximum of the magneto-resistivity. To tune SSE in resonance with MW field, the potential difference $V_{\textrm{B-T}}$ between the bottom and the top electrodes was swept in such a way as to keep the potential at the helium surface to be constant and equal to +10~V. In addition, a constant negative voltage of -30~V was applied to guard rings to make a sharper density profile at the electron sheet edge. The absorption line shape was obtained by recording the bolometer signal as voltage was swept through the resonance. Like in Ref.~4, the signal proportional to $dP_{\textrm{A}}/dE_{\perp}$, the derivative of power absorption with respect to the pressing electric field, was obtained by applying small modulation to $V_{\textrm{B-T}}$. Examples of experimental traces obtained using low MW input power and taken at T=0.55, 0.65 and 0.73~K are shown in Fig.~\ref{fig:deriv}.

\subsection{Results and discussion} 

\begin{figure}[t]
\centering
\includegraphics[width=11cm]{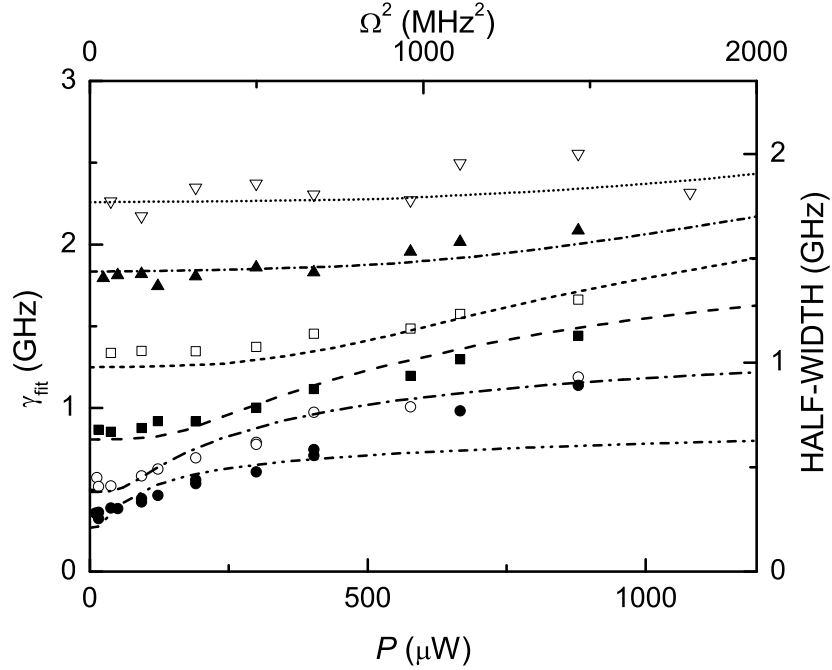}
\caption{\label{fig:width} Experimental linewidth $\gamma_{\textrm{fit}}$ plotted as a function of input MW power $P$  and calculated half-width plotted as a function of $\Omega^2$ for $T$=0.5~K ($\bullet$ and dash-double-dotted line), 0.55~K ($\circ$ and dash-dotted line), 0.6~K ($\blacksquare$ and dashed line), 0.65~K ($\square$ and short-dashed line), 0.7~K ($\blacktriangle$ and short-dash-dotted line) and 0.73~K ($\triangledown$ and dotted line).}
\end{figure}

To see the effect of heating on the absorption linewidth and to make comparison with theory, the experimental traces were recorded at different values of the input MW power $P$ and the linewidth $\gamma_{\textrm{fit}}$ was obtained by fitting the traces to the analytical formula using $\gamma_{\textrm{fit}}$ as an adjustable parameter. To account for line asymmetry and to obtain better fitting in the whole temperature and input power ranges, we employed the fitting with an asymmetric Fano-like line given by~\cite{Isshiki}

\begin{equation}
F(\omega_{12})=A\frac{(\omega_{12} - \omega -B\gamma_{\textrm{fit}})^2}{(\omega_{12} -\omega)^2+\gamma_{\textrm{fit}} ^2},
\label{eq:Fano}
\end{equation} 

\noindent with $A$, $B$, $\omega$ and $\gamma_{\textrm{fit}}$ being fitting parameters. Here $A$ is the amplitude of the resonance, $B$ is a parameter which results in the asymmetry and $\omega$ is the MW frequency, which was allowed to vary to account for the frequency shift discussed in Sec.~3.2. Examples of such fitting are shown as dotted lines in Fig.~\ref{fig:deriv}. At high temperatures and low values of $P$, the experimental traces could be fitted well with a single asymmetric line given by equation (\ref{eq:Fano}). However, at high power the line shape broadened in such a way that it was no longer possible to fit it satisfactorily with a single line. In this case, the experimental data were fitted with the sum of two lines (\ref{eq:Fano}) sharing the same parameter $\gamma_{\textrm{fit}}$. The power dependence of $\gamma_{\textrm{fit}}$ extracted from fitting of experimental traces taken at different values of $T$ are plotted as symbols in Fig.~\ref{fig:width}. The conversion factor to frequency of 2.03~GHz/V was found experimentally by changing the frequency of microwaves and observing the shift of the resonance line. To compare with hot electron theory, in the same figure we plot the half-width of the power absorption curves calculated using (\ref{eq:abs}). The half-width is shown by lines and plotted as a function of $\Omega^2$. Like in the case of comparison between experimental and theoretical resistivity change described earlier in this paper, the horizontal axis for theoretical curves was adjusted to give the best fit between experimental results and theoretical curves. Both methods give comparable estimates for $\Omega$. The values of $\Omega$ for microwaves in the cell can be also estimated from the magnitude of the power absorption signal at InSb bolometer. These values are also consistent with above estimate. The absolute value of experimental linewidth turned out to be only about 30$\%$ higher than theoretically calculated. This deviation from the theory is significantly smaller then previously reported.~\cite{Isshiki} A number of factors can account for this discrepancy. Following Ando~\cite{Ando}, in previous calculations of theoretical linewidth $\gamma$ we used approximate variational method to determine eigen functions of the ground and the first excited states. This gave the linewidth to be about 15$\%$ lower than obtained in the present calculations using numerical solutions of Schrodinger equation for electron eigen functions. Also, the conversion factor, which is used to express the experimental linewidth in frequency units, was previously obtained from calculations and turned out to be about 10$\%$ lower than the one that was experimentally determined and used in the present work. The rest can be probably accounted by better fitting procedure and by improved alighnment of the experimental setup.   

In general, we find rather good agreement between experimentally observed and theoretically expected behaviours. At low temperatures ($T\leq 0.5$~K), the heating becomes significant even at lowest power, and linewidth rises quickly with $P$. At intermediate temperatures ($T=0.5-0.65$~K), the heating is not strong at small values of $P$, and linewidth have weak power dependence. As the power increases, the heating becomes appreciable, and the line broadens. At high temperatures ($T\geq 0.7$~K), the heating is negligible for all values of $P$, and the linewidth is almost power independent. The deviation between theory and experiment was observed only for data taken at $T$=0.5~K and $P\gtrsim500~\mu$W. However, we found out that at $T\leq 0.5$~K and high MW power a different phenomenon appeared. Under these conditions, the line shape became more asymmetric, and the integrated absorption line showed an offset. In addition, at $T\leq 0.3$~K a quadrature component of the modulated absorption signal was recorded. Very similar effect was also observed by Glasson $\textit{et~al}$ in their MW absorption experiment~\cite{Glasson} with SSE over liquid $^4$He and was attributed to the absorption hysteresis due to Coulomb interaction between electrons. It is not completely clear at the moment what role does electron heating play in hysteresis, and the further discussion of this effect is beyond the scope of the present report.

\section{Conclusions}

In conclusion, we observed the resonant increase in the electron resistivity upon irradiation of the SSE with resonant MW radiation. This effect is caused by the heating of SSE with the absorbed MW power and is due to the increase of the inter-subband scattering rate of the hot electrons. It is also demonstrated, both theoretically and experimentally, that the electron heating results in the broadening of the absorption line even long before the absorption saturation can be reached. Our results indicate that under the typical conditions of the MW absorption experiment, the electron heating can not be ignored, and the proper account for this effect is necessary for the adequate analysis of the experimental results.

It would be interesting to observe the resistivity change due to electron heating at low $T$ where the momentum relaxation of the SSE is limited by the interaction with ripplons. Unlike in a vapor-atom scattering regime, in this case the inter-subband scattering rates decrease rapidly with increasing subband index. In addition to this, at high pressing fields the intra-subband relaxation rates also decrease with increasing $T_\textrm{e}$. Therefore, $\sigma^{-1}$ is expected to decrease rapidly with $T_\textrm{e}$. A new experiment to investigate this effect is currently in progress. In addition, a model, which adequately takes into account the electron-ripplon interaction, have been developing. 

\section*{Acknowledgement}

The work is partly supported by the Grant-in-Aids for Scientific Research from Monka-sho and JSPS. One of the authors (D.K.) thanks JSPS for a postdoctoral fellowship. We appreciate valuable discussion with M. Dykman.


\begin{thebibliography}{99}
\bibitem{shift} The image charges contribute to the pressing field if the liquid level is not precisely in the middle between the plates. From the shift of the position of the resonance, the liquid level is estimated to be about 0.05~mm below the middle point.
\end{thebibliography}
\end{document}